\documentclass{emulateapj} % 
\IfFileExists{srcltx.sty}{\usepackage[active]{srcltx}}

\usepackage{apjfonts}

\usepackage{amssymb}% 
\usepackage{amsmath}% 
\usepackage{graphicx}%
\newcommand{\eprint}[1]{\href{http://arxiv.org/abs/#1}{#1}}

\newcommand{\adsurl}[1]{\href{#1}{ADS}}
\providecommand{\url}[1]{\href{#1}{#1}}

\newcommand{\cluster}{\mbox{1E\,0657--56}} %
\newcommand{\chandra}{\emph{Chandra}}

\newcommand{\fov}{{\mathrm{fov}}} % Field-of-view
\newcommand{\xmm}{{\textsc{xmm}}} % XMM 
\newcommand{\dm}{{\textsc{dm}}} % Dark Matter
\newcommand{\cm}{\:\mathrm{cm}} % keV
\newcommand{\ev}{\:\mathrm{eV}} % keV
\newcommand{\kev}{{\:\mathrm{keV}}} % keV
\newcommand{\mpc}{\:\mathrm{Mpc}} % Mpc
\newcommand{\gpc}{\:\mathrm{Gpc}} % Gpc
\newcommand{\parfrac}[2]{\left(\frac{#1}{#2}\right)}

\begin{document}
\title{\hfill\vbox{\scriptsize\hbox{CERN-PH-TH/2006-226}}\hfill
  \\[\bigskipamount]
  Constraints on parameters of radiatively decaying dark matter from
 the galaxy cluster 1E\,0657--56}%

\author{A.~Boyarsky\altaffilmark{1,2,3},
O.~Ruchayskiy\altaffilmark{4},
and M.~Markevitch\altaffilmark{5,6}}

\altaffiltext{1}{CERN, PH-TH, CH-1211 Geneve 23, Switzerland} %

\altaffiltext{2}{\'Ecole Polytechnique F\'ed\'erale de Lausanne, Institute of
  Theoretical Physics, FSB/ITP/LPPC, BSP 720, CH-1015, Lausanne, Switzerland}

\altaffiltext{3}{On leave from Bogolyubov Institute of Theoretical Physics,
  Kyiv, Ukraine}

\altaffiltext{4}{Institut des Hautes \'Etudes Scientifiques, Bures-sur-Yvette,
 F-91440, France}

\altaffiltext{5}{Harvard-Smithsonian Center for Astrophysics,
 Cambridge, MA 02138, USA} % 

\altaffiltext{6}{Space Research Institute, Moscow, Russia} % 

\setcounter{footnote}{6}

\begin{abstract}
  
  We derived constraints on parameters of a radiatively decaying warm dark
  matter particle, e.g., the mass and mixing angle for a sterile neutrino,
  using \chandra\ X-ray spectra of a galaxy cluster \cluster\ (the ``bullet''
  cluster). The constraints are based on nondetection of the sterile neutrino
  decay emission line.  This cluster exhibits spatial separation between the
  hot intergalactic gas and the dark matter, helping to disentangle their
  X-ray signals. It also has a very long X-ray observation and a total mass
  measured via gravitational lensing. This makes the resulting constraints on
  sterile neutrino complementary to earlier results that used different
  cluster mass estimates.  Our limits are comparable to the best existing
  constraints.

\end{abstract}


%\pacs{95.35.+d, 14.60.Pq, 95.85.Nv} 

%%\maketitle

\section{Sterile neutrino as Warm DM candidates}
\label{sec:introduction}

A number of works appeared recently on the subject of a sterile (or
\emph{right-handed}) neutrino as a possible dark matter (DM) candidate
\citep[e.g.,][]{Asaka:05a, Asaka:05b, Abazajian:05a, Abazajian:05b,
  Boyarsky:05, Boyarsky:06a, Boyarsky:06b, Boyarsky:06c, Boyarsky:06d,
  Asaka:06, Shaposhnikov:06, Riemer:06, Watson:06}.  The discovery of
neutrino oscillations (see, e.g., \citealt{Strumia:06} for a review) made
the existence of a sterile neutrino quite plausible and spurred the interest
in this candidate. Several factors make dark matter made of sterile
neutrinos with a mass in the keV range particularly interesting:

(i) It is the lowest possible range of masses for fermionic DM. The Pauli
exclusion principle applied to the DM particles in the halos of dwarf
spheroidal galaxies (such as Draco or Ursa Minor) implies a lower bound on the
particle mass $M_\dm\gtrsim 350 \ev$~\citep{Tremaine:79}.  Thus the sterile
neutrino can be light enough to be a \emph{warm} DM candidate (see below).

(ii) Warm DM with a keV mass can alleviate the problem of too many small
subhalos inside the bigger dark matter halos, and too sharp central density
peaks in the galaxy-sized DM halos predicted in the Cold Dark Matter
scenario~\citep{Bode:00,Goerdt:06}. For example, the flat central radial
mass profile of the Fornax dwarf spheroidal
galaxy~\citep{Goerdt:06,Strigari:06} can be explained if dark matter is warm
with $M_\dm \sim 2\kev$.
  
(iii) \citet{Asaka:05a} and \citet{Asaka:05b} recently demonstrated that a
simple extension of the Standard Model by three singlet fermions with masses
smaller than the electroweak scale can accommodate the data on neutrino masses
and mixings, provides a candidate dark matter particle (in the form of the
lightest sterile neutrino), and can explain the baryon asymmetry of the
Universe. Such an extension (dubbed $\nu$MSM) can quantitatively explain these
%problems %
``beyond the Standard Model'' phenomena
within a single consistent framework. It should be tested observationally, and
one such test is the search for the sterile neutrino DM.
 
(iv) For a DM particle with the mass in the keV range, one can obtain lower
bounds on its mass by modeling the large scale structure formation.  The
power spectrum of the matter density fluctuations derived from the
Lyman-$\alpha$ forest data in the SDSS, spanning redshifts $2.2<z<4.2$
\citep{Seljak:06,Viel:06} constrains the mass of a warm DM particle to the
range $\gtrsim 10\kev$ ($\gtrsim 14\kev$ in \cite{Seljak:06}). However, a
more conservative analysis using only the higher spectral resolution
Lyman-$\alpha$ data and lower redshifts gives $M_\dm\gtrsim X\kev$
\citep{Hansen:01,Viel:05}.

(v) Sterile neutrinos with a keV mass would have other interesting
astrophysical applications
\citep[e.g.,][]{Kusenko:06a,Kusenko:06b,Biermann:06,Stasielak:06}.

%%%%%%%%%%%%%%%%%%%%%%%%%%%%%%%%%%
\begin{figure*}
% \newcounter{widefigure} 
% \setcounter{widefigure}{figure}
% \renewcommand{\thefigure}{\thewidefigure.\alph{figure}}
%  \begin{minipage}[c]{.45\textwidth}
%   \includegraphics[width=\textwidth]{xray_submain_r6_ed} %
%   \label{fig:sub_main} \end{minipage} &
% \begin{minipage}[c]{.45\textwidth}
%    \includegraphics[width=\textwidth]{lens_submain_r6_ed} % 
%\end{minipage}
%\end{tabular}
%
\plottwo{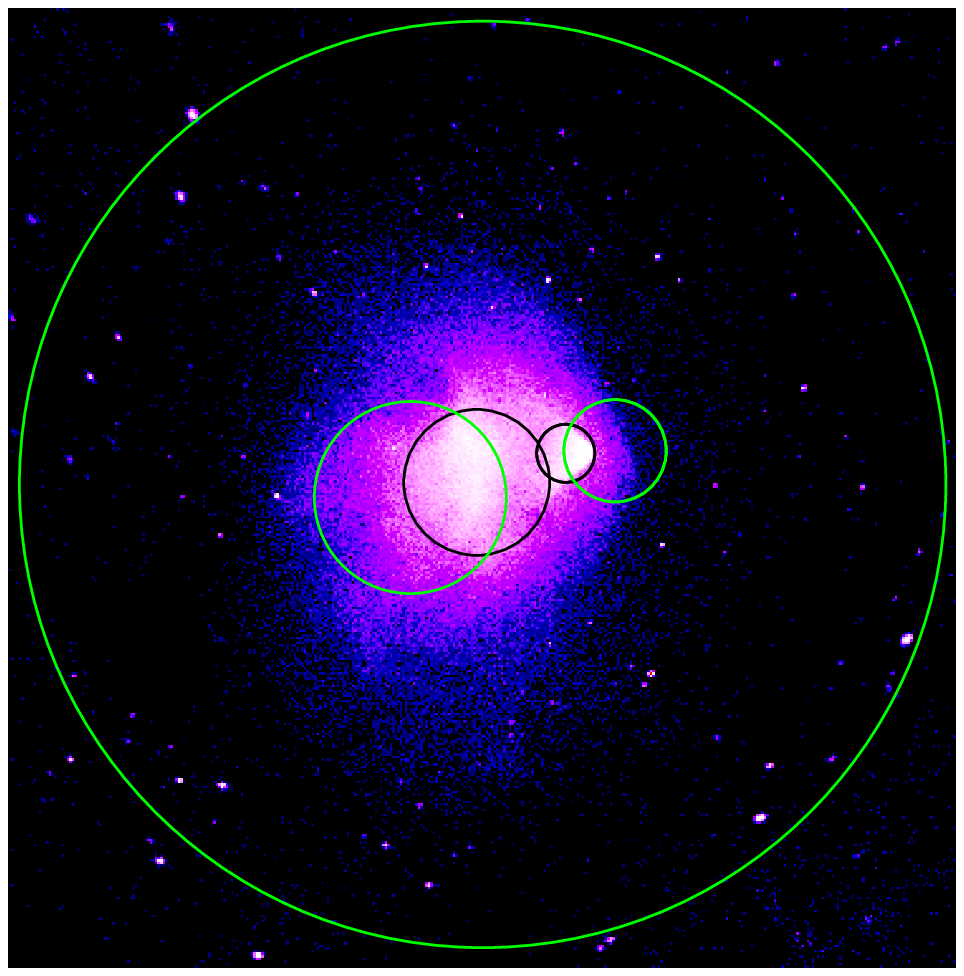}{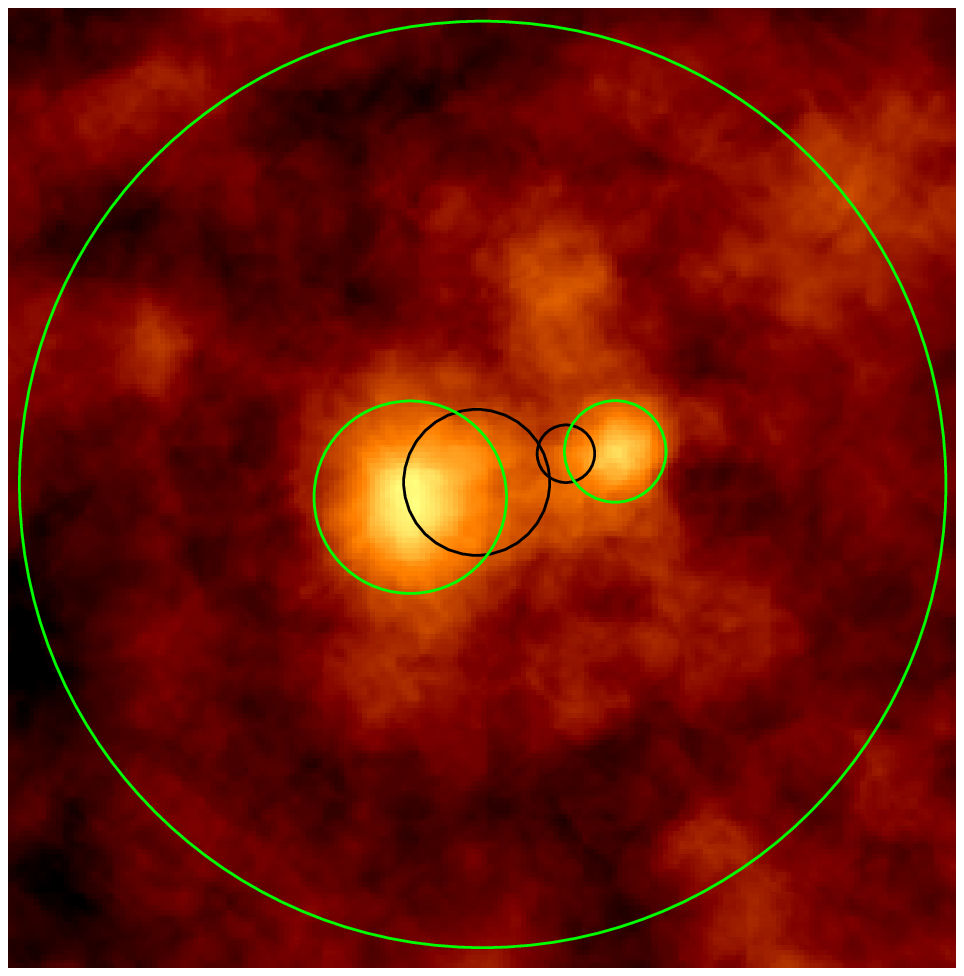}
\label{fig:r3m}
\caption{Regions used for extracting the X-ray spectra overlaid on a
  \chandra\ X-ray 0.8--4 keV image of \cluster\ ({\em left panel})\/ and its
  weak lensing mass map \citep{Clowe:06a} ({\em right panel}).  The panel
  size is $12'\times 12'$. All point sources seen in the X-ray image are
  spatially excluded from the spectral analysis.  The region that we refer
  to as {\sc sub} is the green circle centered on the western subcluster's
  mass peak, excluding the smaller black circle encompassing most of the
  X-ray gas bullet.  The region {\sc peaks} is a combination of {\sc sub}
  and the green circle centered on the eastern mass peak, again excluding
  the corresponding X-ray peak (the bigger black circle).  The region {\sc
    whole} is the big ($r=6'$) green circle, excluding only the gas bullet
  (the smaller black circle). \vspace{3mm}}
\label{fig:regions}
\end{figure*}
\vspace{5mm}
%%%%%%%%%%%%%%%%%%%%%%%%%%%%%%%%%%

The sterile neutrino should possess a radiative decay channel
\citep{Pal:81,Barger:95} (see \S\ref{sec:decay} below). An emission line
from neutrino decay has been searched for --- so far unsuccessfully --- in
the X-ray spectra of various types of astrophysical objects, including
clusters of galaxies~\citep{Abazajian:01b,Boyarsky:06b}, the diffuse X-ray
background~\citep{Dolgov:00,Boyarsky:05}, the DM halo of the Milky
Way~\citep{Boyarsky:06c,Boyarsky:06d,Riemer:06}, dwarf galaxies
\citep{Boyarsky:06c,Boyarsky:06d}, and the M31 galaxy~\citep{Watson:06}.
Assuming that all of the dark matter is in the form of sterile neutrinos,
nondetection of such af line places constraints on the mixing angle of the
sterile neutrino as a function of mass in the range $1\kev \lesssim
M_s\lesssim 100\kev$. To derive such constraints, one needs to know the mass
of the DM in the field of view of the X-ray spectrometer, $M_\dm^\fov$
(eq.~\ref{eq:2} below). There are several ways of deriving these masses:

(a) Modeling rotational curves of stars in galaxies or the velocity
dispersion of galaxies in dynamically relaxed galaxy clusters;

(b) reconstructing the density and temperature profiles of the hot
intergalactic gas in relaxed galaxy clusters using X-ray data and determining
the total mass from the assumption of hydrostatic equilibrium
\citep[e.g.,][]{Cavaliere:76,Bahcall:77,Vikhlinin:05};

% THE CORRECT REF. FOR SARAZIN:77 IS
% Bahcall, J. N., \& Sarazin, C. L. 1977, ApJ, 213, L99

(c) For sufficiently distant ($z>0.1$) clusters, gravitational lensing can
be used \citep{Bartelmann:01}, which does not require a cluster to be
relaxed.

All previous observational constraints
%\citep{Abazajian:01b,Boyarsky:05,Boyarsky:06b,Boyarsky:06c,Riemer:06,
%  Watson:06,Boyarsky:06d} 
were derived for nearby objects ($z\lesssim 0.01$), for which a combination of
methods (a) and (b) provided the DM density distribution and $M_\dm^\fov$. It
is important to cross-check these results using objects with masses determined
by all methods, to minimize the systematic uncertainties inherent in each
method.

In the present work, we use a distant object ($z=0.296$, corresponding to
the luminosity distance $D_L=1.530\gpc$ for our adopted cosmology with
$h=0.7$, $\Omega_\mathrm{m}=0.3$ and $\Omega_\Lambda = 0.7$), whose mass is
determined via weak and strong gravitational lensing
\citep{Clowe:06a,Bradac:06}. This method gives the total projected mass
within a given region in the sky, eliminating a number of uncertainties of
other methods. It is the only mass measurement method applicable to this
cluster, which undergoes a violent merger. The merger has also resulted in a
unique separation between the dark and visible matter \citep{Clowe:06a},
which we will utilize below.

\subsection{Radiative decay of sterile neutrinos}
\label{sec:decay}

A dark matter composed of sterile neutrinos should not be completely
``dark''~\citep{Dolgov:00,Abazajian:01b}.  The sterile neutrino possesses a
radiative decay channel, decaying at a rate $\Gamma$ into an active neutrino
and a photon with energy $E=M_s/2$ (where $M_s$ is the sterile neutrino
mass). It is convenient to parameterize the interaction with the active
neutrino in terms of the mixing angle $\sin^2 (2\theta)$.  The decay rate
$\Gamma$ is then given by~\citep{Pal:81,Barger:95}:
\begin{align}
 \label{eq:1}
  \Gamma & = \frac{9\, \alpha\, G_F^2}{1024\pi^4}\sin^2 (2\theta)\, M_s^5 \notag\\
  & =  1.38\times10^{-22} \sin^2 (2\theta)
 \left[\frac{M_s}{1\,\mathrm{keV}}\right]^5 \,\mathrm{s^{-1}}.
\end{align}

For distant objects, the decay flux into a solid angle $\Omega_\fov$ (the
spectrometer's field of view, FoV) is given by
\begin{align}
 \label{eq:2}
 F_\dm = \frac{M_\dm^\fov \Gamma}{4\pi D_L^2}\frac{E}{M_s}
\end{align}
where $M_\dm^\fov$ is the total mass of DM within this solid angle. If the
object is at a redshift $z$,
\begin{align}
\label{eq:3}
F_\dm &=6.4\parfrac{M_\dm^\fov}{10^{14}M_\odot}\parfrac{100\mpc}{D_L}^2\notag\\
&\times\sin^2 (2\theta)
\left[\frac{M_s}{\mathrm{1 keV}}\right]^5 \kev\; \mathrm{cm^{-2}\; s^{-1}}.
\end{align}
Nondetection of any X-ray emission lines in the spectrum of a massive object
that are not expected from its baryonic constituents (the hot intergalactic
gas in the case of a galaxy cluster) can be used to place an upper limit on
the flux from sterile neutrino decay. Eq.~(\ref{eq:3}) can then be used to
constrain the parameters $M_s$ and $\theta$.

\section{The \cluster\ cluster}
\label{sec:1e-cluster}

\cluster\ is an interesting object for constraining the brightness of the
neutrino line, for several reasons. Its total mass is directly measured from
gravitational lensing \citep{Clowe:06a,Bradac:06}. It also has a very long
(450 ks) \chandra\ observation \citep{Markevitch:05}, which provides a
high-statistic X-ray spectrum. While formal constraints on the neutrino model
that we will obtain below (\S\ref{sec:results}) are not significantly better
than those previously derived from the nearby X-ray clusters Coma and
Virgo~\citep{Boyarsky:06b},
%% REF TO ABAZAJIAN
\cluster\ provides a significant improvement in reliability, because of its
directly measured total mass (whereas Coma and Virgo are both nearby
unrelaxed systems, so their dark matter masses, and how much of it falls
inside the instrument FoV, is uncertain).

A unique feature of \cluster\ is a spatial separation between the peaks of
gas and dark matter density belonging to the two subclusters
\citep{Clowe:06a}, caused by their merger in the plane of the sky (Fig.\ 
\ref{fig:regions}). This enables us to try to exclude the spatial regions
with the highest thermal X-ray contamination (and thus minimize the
uncertainty of modeling this component, see \S\ref{sec:uncertainties}
below), while at the same time retaining the densest dark matter regions.
Thus, we will use two regions in our X-ray analysis below, shown in
Fig.~\ref{fig:regions}.  The region {\sc peaks} takes advantage of the
separation between the DM and gas and combines two circles centered on the
mass peaks, excluding the two X-ray brightness peaks. We will also use a
subregion of {\sc peaks} that includes only the bullet subcluster mass peak,
{\sc sub}, to illustrate the effects of uncertainties. The region {\sc
  whole} includes most of the cluster mass within $r=6'=1.6$ Mpc (a still
bigger region will increase uncertainties of the mass and the detector
background), excluding only a small region at the X-ray brightness peak.

%%%%%%%%%%%%%%%%%%%%%%%%%%%%%%
\begin{table}[t]
\centering
\caption{Masses within spectral extraction regions}
\begin{tabular}{lcccc}
\hline
\hline
  Region & Total Mass,$^a$  & Gas mass,   & DM mass, & area,$^b$ \\
     & $10^{15} M_\odot$& $10^{15} M_\odot$& $10^{15} M_\odot$ &
     $10^{-7}$ sr\\
\hline
  {\sc Sub} & 0.058 & 0.007 & 0.05 & 1.00\\
  {\sc Peaks} & 0.198 & 0.034 & 0.16 & 3.55\\
  {\sc Whole} & 1.46 & 0.297 & 1.16 & 95.3\\
%  \texttt{Main} & 0.14 & 0.027 & 0.11 & 2.55\\
%  \texttt{r3m} & 0.75 & 0.191 & 0.56 & 23.5\\
\hline

\multicolumn{5}{l}{$^a$ Masses from weak lensing}\\
\multicolumn{5}{l}{$^b$ The solid angle of the region}\\
\end{tabular}
\label{tab:masses}
\end{table}
%%%%%%%%%%%%%%%%%%%%%%%%%%%%%%

To calculate the dark matter mass within the spectral extraction regions, we
integrated the weak lensing map of the projected total
mass~\citep{Clowe:06a} and subtracted a relatively small contribution from
the intracluster gas. We note that the mass near the cluster center derived
from weak lensing is lower by a factor of about 2 compared to that derived
by Brada\v{c} et al., who combined weak and strong lensing data (but whose
fit is in fact dominated by the strong lensing data).  Some of this
discrepancy is expected, because the weak lensing approximation breaks down
near the peaks of massive clusters that produce strong gravitational arcs
(such as both subclusters of \cluster); indeed, the peak densities in
\cite{Clowe:06a} are insufficient to produce arcs.  Weak lensing is also
insensitive to adding a constant mass sheet \citep{Bartelmann:01}.  Strong
lensing analysis may suffer from other types of uncertainties. The
higher-mass \citet{Bradac:06} map is limited to the central $r=1.5'-2'$
region, insufficient for our purposes, so we chose to use the
\cite{Clowe:06a} mass map. This will result in conservative underestimates
of the expected neutrino signal by a factor of up to 2. We will illustrate
this uncertainty in the final results.

The gas mass within our spectral regions is estimated from a
three-dimensional model fit to the \chandra\ X-ray brightness and
temperature maps. The X-ray emissivity of a hot gas ($T\sim 8-20$ keV) in
the \chandra\ energy band is determined mostly by the gas density, and
depends weakly on temperature. Clusters are optically thin for X-rays, so
in general, their gas density can be reconstructed very reliably for
symmetric clusters. The apparent axial symmetry of \cluster\ allowed us to
derive a gas model with a 10\% accuracy~\cite{Markevitch:06}. The gas
contribution to the total mass in our spectral regions is about 25\% (for
the total mass from weak lensing), which we subtracted to obtain the dark
matter masses given in Table \ref{tab:masses}. Given the small contribution
of gas to the total mass, its uncertainty will be neglected.

%%%%%%%%%%%%%%%%%%%%%%%%%%%%%%%%%%
\begin{figure}[b]
\vspace{7mm}
\centering

\includegraphics[height=0.95\columnwidth,angle=-90,bb=34 0 502 671,clip]%
{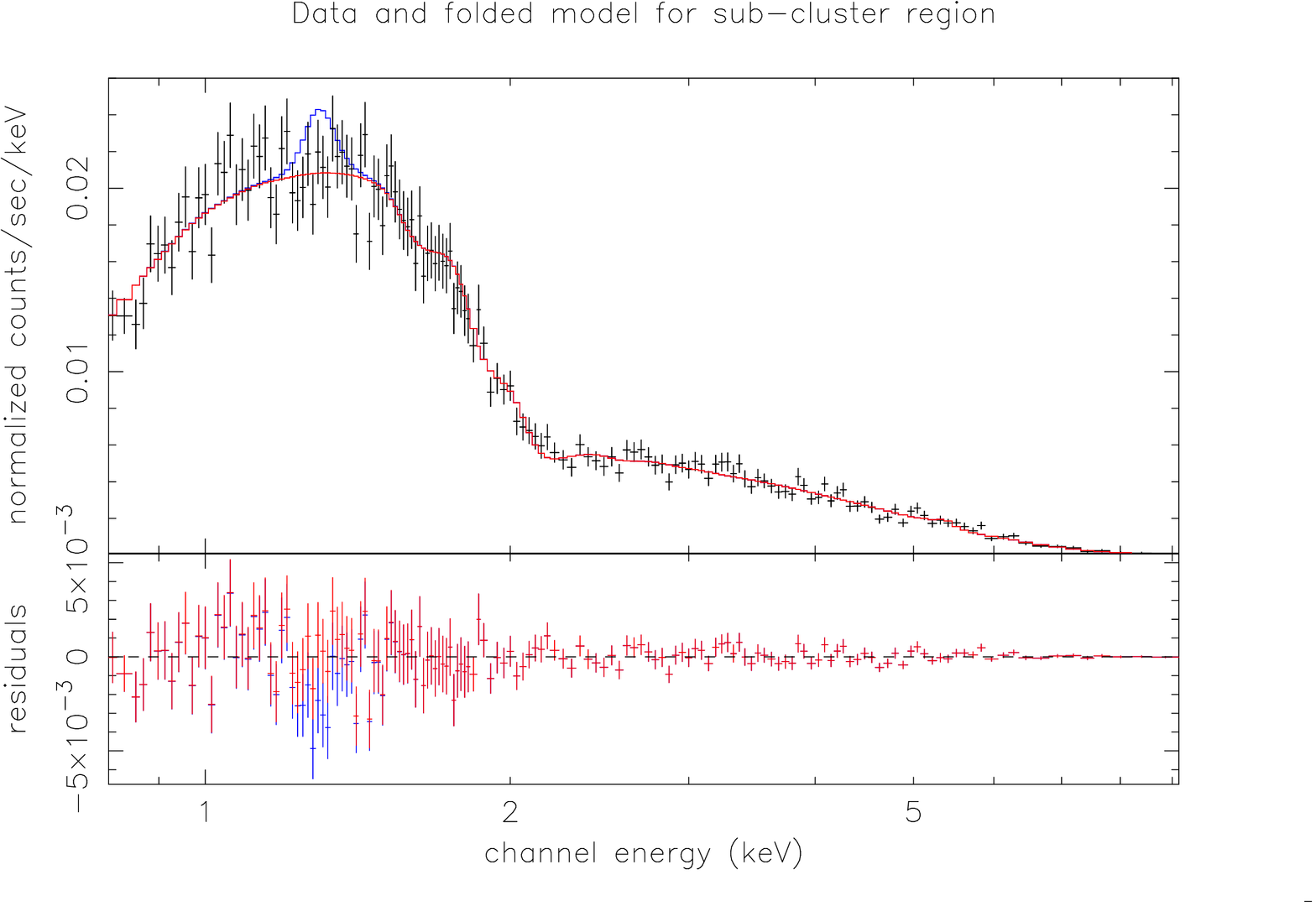}
 \caption{Spectrum for the {\sc sub} region with the best-fit {\sc apec}
   model showin in red.  For illustration, we show an additional narrow line
   at $E=1.3\kev$ (blue model line and residuals), which worsens the fit at
   a $3\sigma$ level.}
 \label{fig:sub-xspec}
\end{figure}
%%%%%%%%%%%%%%%%%%%%%%%%%%%%%%%%%%

%%%%%%%%%%%%%%%%%%%%%%%%%%%%%%%%%%
\begin{figure}[t]
\centering
\includegraphics[width=\columnwidth]{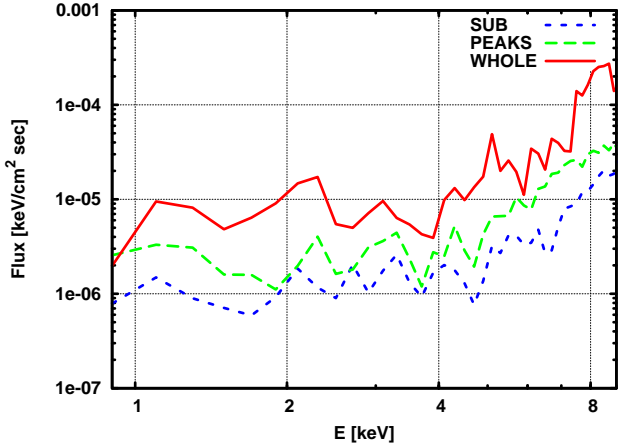}
  \caption{Statistical upper limits ($3\sigma$) for the flux in a
    nonthermal, narrow emission line as a function of line energy, for our
    fitting regions. \vspace{3mm}}
  \label{fig:fluxes-regions}
\end{figure}
%%%%%%%%%%%%%%%%%%%%%%%%%%%%%%%%%%

\section{Data analysis}
\label{sec:data-analysis}

We use the deep \chandra\ \citep{Weisskopf:02} ACIS-I observation of \cluster\ 
described in \cite{Markevitch:05,Markevitch:06}. The ACIS non-grating energy
resolution is between 12\%--4\% (half-power line width $\Delta E/E$) in the
$E=1-8$ keV range. The X-ray spectra, the blank-sky background spectra, and
ACIS responses (that include all instrument effects such as mirror effective
area, detector efficiency and energy resolution, to be applied to a model
spectrum in order to compare it to the data) were derived as described in
\cite{Markevitch:06} and \cite{Vikhlinin:05}.  Spectral fitting was performed
using the {\sc xspec} package \citep{Arnaud:96}.  The spectra for our three
regions, {\sc sub}, {\sc peaks}, and {\sc whole}, were well-fit (with reduced
$\chi^2\approx 1$ in all cases) with models consisting of one or several
thermal plasma components ({\sc apec}, \citealt{apec}) representing the
multitemperature intracluster gas.  A useful property of thermal spectra is
that a continuous range of gas temperatures produces a spectrum that can be
adequately fit with a sum of just several discrete temperatures. Models for
all regions were modified at low energies by the Galactic absorption with $N_H
= 4.6\times 10^{20}\cm^{-2}$. An example of the fit for the region {\sc sub}
is shown in Fig.~\ref{fig:sub-xspec}; this fit has a reduced $\chi^2=0.98$ for
152 d.o.f.

None of the regions exhibits any emission lines other than those expected
from hot gas (mostly the $E\simeq 6.7$ keV Fe line, redshifted to 5.2 keV).
We use this fact to place constraints on the neutrino decay flux following
the procedure previously used in this context by, e.g.,
\cite{Boyarsky:05,Boyarsky:06b,Boyarsky:06c,Boyarsky:06d}. In particular,
for each energy bin in the 0.8--9 keV range, we added a narrow emission line
(a Gaussian line much narrower than the detector spectral resolution) to the
thermal model, re-fit the spectra, and calculated a statistical upper limit
on the line flux by increasing the line normalization until $\chi^2$ of the
fit worsens by %4 ($2\sigma$ or 95\% confidence level) or 
9 ($3\sigma$ or 99.7\% confidence level). To obtain conservative upper limits,
we allowed as much freedom for the parameters of thermal model as possible,
including allowing the heavy element abundances (that produce the thermal
emission lines) to vary, thus letting the neutrino line mimic some of the
thermal line flux at the respective energies. We note that the full number of
counts in each bin (including the instrumental background) is sufficiently
high to ensure the Gaussian statistics, and is much higher than the resulting
limits on the line flux, so the use of $\Delta \chi^2$ is appropriate
\citep[cf.][]{Protassov:02}. An example of an emission line that would
correspond to a $3\sigma$ upper limit is shown in Fig.\ \ref{fig:sub-xspec},
and the resulting statistical limits, in Fig.\ \ref{fig:fluxes-regions}.

\subsection{Systematic uncertainties}
\label{sec:uncertainties}

\begin{figure}

\epsscale{1.1}
\plotone{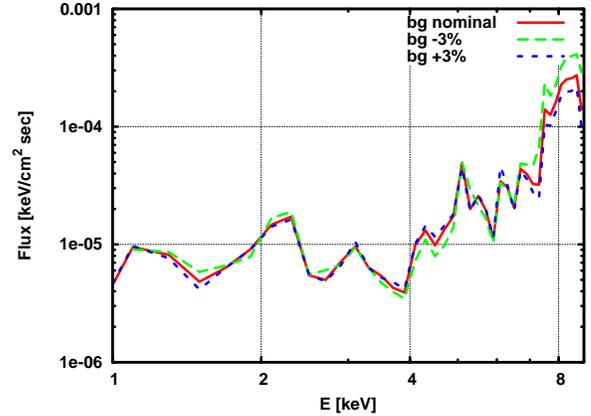}

\caption{The effect of a $\pm3$\% systematic uncertainty in the ACIS background
  normalization for the region {\sc whole} on our line limits.  It is
  significant for large regions and at high energies; for smaller regions such
  as {\sc sub} and {\sc peaks}, it is negligible (not shown).  \vspace{3mm}}
 \label{fig:r6m-bg}
\end{figure}

In addition to the statistical upper limits on the line flux, there are
systematic uncertainties that has to be taken into account.  First, the way
how we normalize the ACIS background using the high-energy band
\cite{Markevitch:03,Hickox:06} results in a 3\% uncertainty of the
normalization at the useful energies. Because the ACIS detector background
has several prominent emission lines, such incorrect normalization may, for
example, hide an emission line coming from the sky, or create a spurions
line.  To take this into account, we varied the background normalization by
$3\%$ and repeated the fitting procedure. As expected, this leads to a
noticeable increase of the allowed line flux only at rather high energies
$E\gtrsim 6\kev$, where the background intensity increases steeply. For the
region {\sc whole}, limits with nominal background normalization and those
of for normalization changed by $\pm 3\%$ are shown in
Fig.~\ref{fig:r6m-bg}.  These differences were added in quadrature to the
statistical limits at each energy.

\begin{figure}[b]
\vspace{5mm}
\centering %

\includegraphics[height=0.95\columnwidth,angle=-90,bb=112 44 556 706,clip]%
{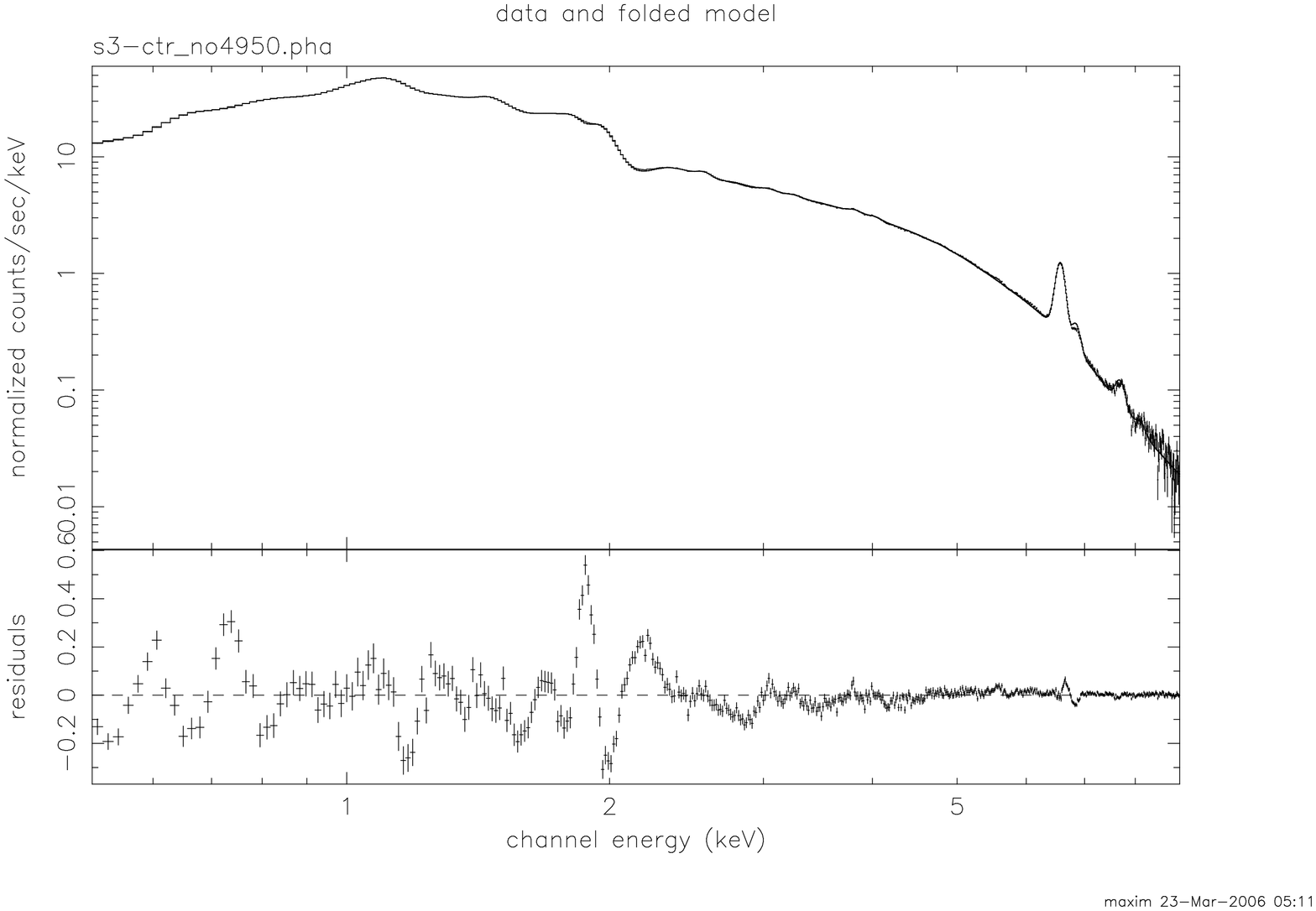}

\caption{\chandra\ ACIS spectrum of the Perseus cluster from
  a 880 ks exposure, extracted from the $8'\times 8'$ central region,
  excluding the very center ($r<1'$) with complex gas structure. The fit
  residuals illustrate the current calibration uncertainties. The residuals
  around 2--4 keV are 2--3\%. \vspace{5mm}}
\label{fig:per}
\end{figure}

\begin{figure}
  
  \epsscale{1.1} \plotone{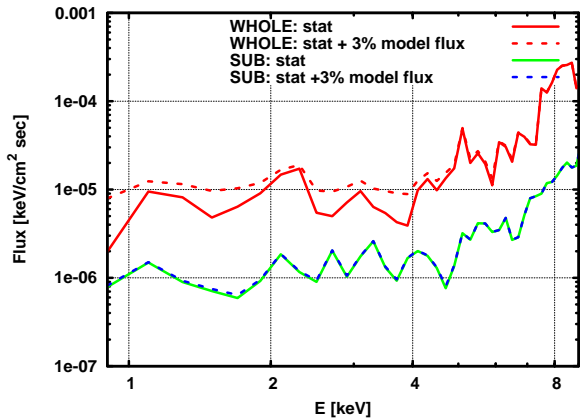}

\caption{The effect of a 3\% calibration uncertainty in the model flux (that
  can take the shape of spurious lines, see Fig.\ \ref{fig:per}). It is
  significant for the line limits derived in regions with bright underlying
  X-ray emission ({\sc whole}), and negligible for regions avoiding the X-ray
  brightness peaks, such as ({\sc sub}).}
 \label{fig:r6m-cal}
\end{figure}

There is a more insidious uncertainty arising from the inaccuracies of
calibration of the detector response and gain (the energy to spectral
channel conversion). To assess this uncertainty, we have extracted a
spectrum from the 880 ks ACIS observation of the the very bright Perseus
cluster, and fit it using the same calibration products (current as of
summer 2006) and a model consisting of several thermal models as we use in
this work, with all element abundances allowed to vary. The fit is shown in
Fig.~\ref{fig:per}. Statistical errors in this dataset are mostly
negligible.  The fit shows systematic residuals at a 2--3\% level of the
model flux, some edge-like or even line-like, obviously caused by
calibration inaccuracies (e.g., the feature around $E=2$ keV is obviously
due to a gain error). To take this uncertainty into account, we added 3\% of
the thermal model flux contained within the width of a Gaussian line, in
quadrature to the statistical limits on the line flux. The result of adding
these uncertainties for regions {\sc whole} and {\sc sub} is shown in
Fig.~\ref{fig:r6m-cal}. As expected, this uncertainty contributes mostly at
lower energies, where thermal emission is bright (at high energies, the
increasing statistical uncertainty starts do dominate).  This is the
uncertainty that can be minimized by observing ``dark'' matter clumps, such
as our gas-stripped subcluster.

\begin{figure*}[t]
\centering
% \begin{tabular}[t]{cc}
%  \begin{minipage}[c]{.5\textwidth}
%   \includegraphics[width=\textwidth]{results-weak-lensing} %
% \end{minipage} &
% \begin{minipage}[c]{.5\textwidth}
%    \includegraphics[width=\textwidth]{r6m-various-limits} % 
% \end{minipage}
% \end{tabular} 

\epsscale{1.1}
\plottwo{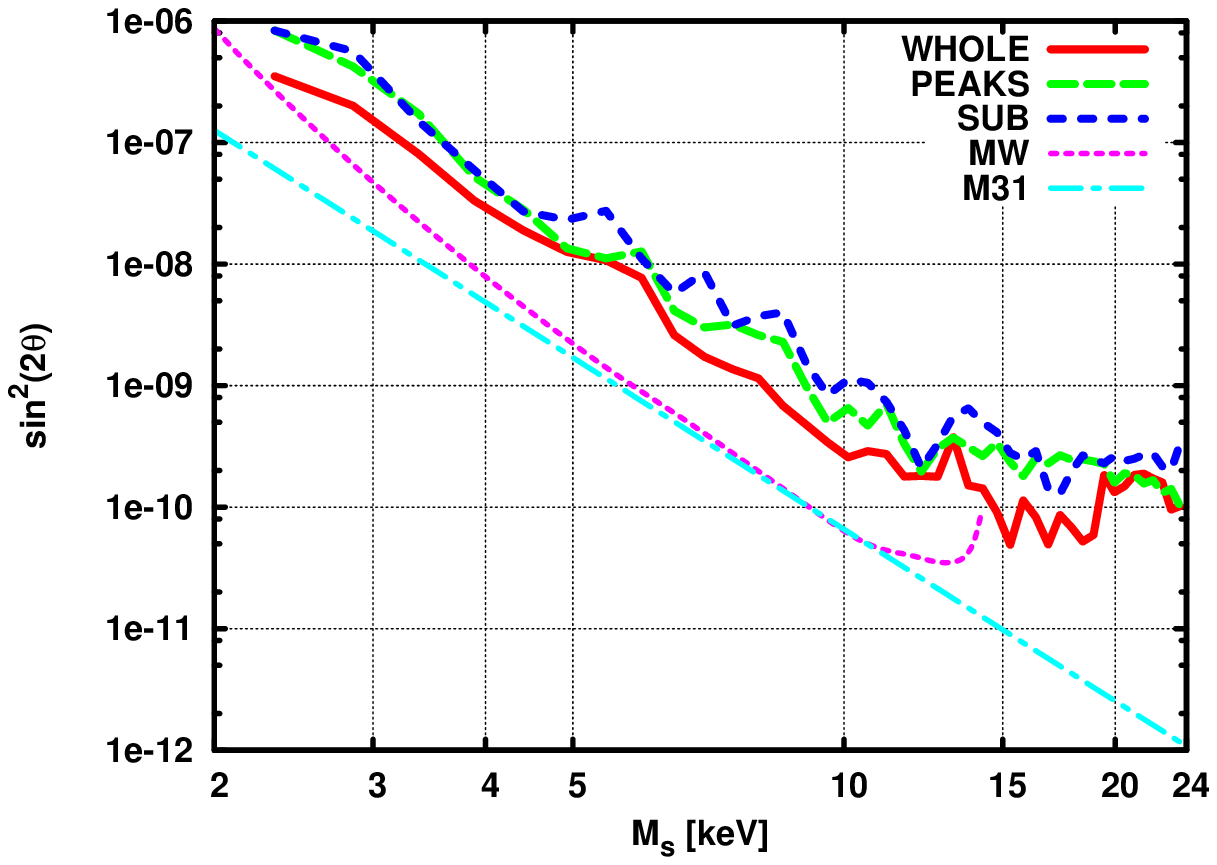}{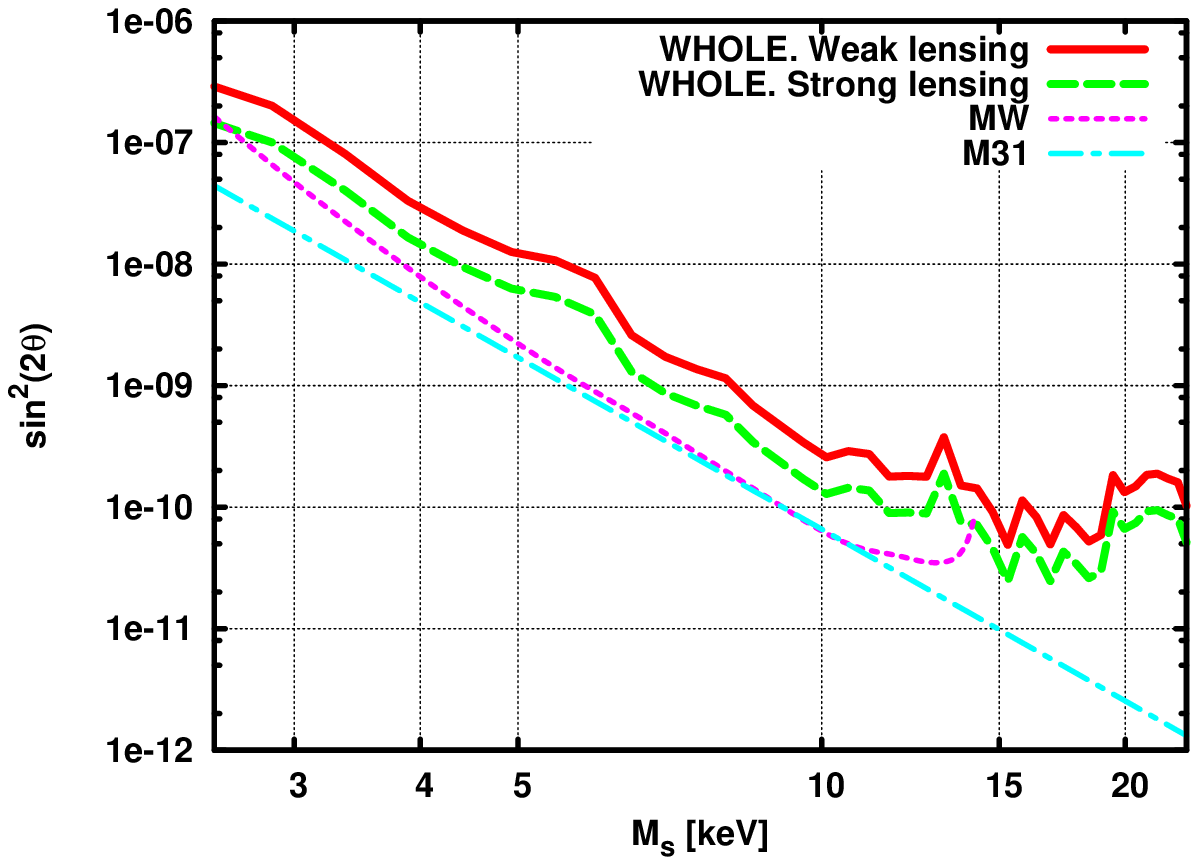}

\caption{({\em a}) Our $3\sigma$ limits on sterile neutrino parameters from
  different spectral regions, using the dark matter masses from weak lensing.
  The area above the curves is excluded.  For comparison, \xmm\ constraints
  from the Milky Way \citep{Boyarsky:06c} and M31~\citep{Watson:06} are shown.
  ({\em b}) Our constraints for region {\sc whole} using masses determined
  from weak lensing (same as in panel {\em a}) and strong lensing.
  \vspace{5mm}}
\label{fig:results}
\end{figure*}

Finally, the biggest uncertainty, unrelated to the X-ray data, comes from
the factor of 2 difference between the cluster masses determined from the
weak and strong gravitational lensing analyses \cite{Clowe:06a,Bradac:06}.
We will include it in the plot with results below
(Fig.~\ref{fig:results}{\em b}).

%%%%%%%%%%%%%%%%%%%%%%%%%%%%%%%%%%
%%%%%%%%%%%%%%%%%%%%%%%%%%%%%%%%%%

%%%%%%%%%%%%%%%%%%%%%%%%%%%%%%%%%%
%%%%%%%%%%%%%%%%%%%%%%%%%%%%%%%%%%

%%%%%%%%%%%%%%%%%%%%%%%%%%%%%%%%%%
%%%%%%%%%%%%%%%%%%%%%%%%%%%%%%%%%%

%%%%%%%%%%%%%%%%%%%%%%%%%%%%%%%%%%
%%%%%%%%%%%%%%%%%%%%%%%%%%%%%%%%%%

\section{Results}
\label{sec:results}

Using eq.~(\ref{eq:3}) and masses from Table~\ref{tab:masses} (and, of course,
the assumption that sterile neutrinos account for all of the dark matter), we
convert upper limits on the neutrino line flux for our two regions into
restrictions on sterile neutrino in the $M_s - \sin^2 (2\theta)$ plane. They
are shown in Fig.\ \ref{fig:results}. The strongest constraint comes from the
region {\sc whole} --- the bigger mass of DM within the field of view turns
out to offer a greater advantage than the reduced thermal contamination at the
gas-stripped DM peaks. For the {\sc sub} and {\sc peaks} regions, we plot
constraints for the conservative (perhaps excessively so) DM mass estimate
from weak gravitational lensing (Fig.~\ref{fig:results}{\em a}) along with a
stronger estimate for the higher strong lensing mass
(Fig.~\ref{fig:results}{\em b}).

Although on average these results are about an order of magnitude weaker (in
terms of $\sin^2 [2\theta]$) than other recent limits
\citep[e.g.,][]{Boyarsky:06c,Watson:06}
\footnote{We note that the M31 constraints presented in~\cite{Watson:06} (as
  a straight line in the $M_s-\sin^2 (2\theta)$ plane) should be very
  qualitative at $M_s\gtrsim 10-12$ keV. They must worsen as our limits do,
  since the \xmm\ effective area rapidly declines at the corresponding
  energies, similarly to \chandra's.}
or more recent results~\citep{Boyarsky:06d}, they serve as an important
cross-check. First, they are obtained from an object with $z\sim 0.3$, while
previous
results~\citep{Boyarsky:05,Boyarsky:06b,Boyarsky:06c,Riemer:06,Watson:06} were
obtained for objects with $z\lesssim 0.01$
\footnote{When this work was in final preparation, a
  preprint~\cite{Riemer:06b} appeared, in which a \chandra\ grating spectrum
  of the cluster A1835 ($z\simeq 0.25$) is analyzed. That analysis strongly
  underestimated the effect of the cluster angular extent on energy
  resolution of the grating spectrum, so we do not consider it here.}.
Furthermore, the DM mass was determined via gravitational lensing, a method
not applicable for nearby objects. This is important, as different mass
measurement methods are subject to different uncertainties, and using an
object such as \cluster\ makes the constraints more robust.

\subsection{Dodelson-Wilson scenario}
\label{sec:dw}

Assuming that sterile neutrinos constitute all the DM and that they are
produced in the early Universe via mixing with active neutrinos only, one
should expect a relation between the mass of sterile neutrino and its mixing
angle~\citep{Dodelson:93,Dolgov:00,Abazajian:01a,Abazajian:05a}.  Combined
with observational restrictions on the DM decay emission, this provides an
upper bound on the sterile neutrino mass.

However, as sterile neutrinos do not thermalize in the early Universe, any
such model relies on a number of assumptions (including initial conditions
at temperatures $ \gtrsim 1$~GeV and the absence of entropy
dilution)~\citep{Boyarsky:06c,Asaka:06}. To define the boundary conditions,
the knowledge of some ``beyond the $\nu$MSM'' physics is needed.  For
example, it was shown in~\cite{Shaposhnikov:06} that all of the DM sterile
neutrinos could have been produced by interaction with inflation.  In such a
scenario, the mixing angle can be arbitrarily small, even zero. This means
that the sterile neutrino mass and mixing angle are not necessarily related,
contrary to the assumption used in recent literature
\citep[e.g.,][]{Abazajian:05a,Watson:06}.  Upper limits on the neutrino mass
placed in those works, based on the upper limits on the decay line flux, are
not only model-dependent, but depend strongly on the initial conditions.
Moreover, even assuming ad hock initial conditions implying that there was
no DM at the temperatures $\gtrsim 1$~GeV (which is hardly physically
justified), the correct calculation of the production rate requires
calculation of the non-preturbative QCD contributions, which is still a
subject of discussion in the literature~\citep{Asaka:06,Asaka:06b}.

Nevertheless, for the sake of comparison with other works, the intersection
of our constraints in Fig.\ \ref{fig:results}{\em a} with the $M_s - \sin^2
(2\theta)$ relation obtained in~\cite{Abazajian:05a} for the simplest DW model
(with one sterile neutrino, assuming zero initial conditions and a
particular form of QCD contribution) corresponds to an upper limit
$M_{s}<6.3$ keV.

\acknowledgments

We would like to thank A.Neronov, M.Shaposhnikov and I.Tkachev for useful
discussions. OR was supported in part by the European Research Training
Network contract 005104 ``ForcesUniverse'' and by a \emph{Marie Curie
  International Fellowship} within the $6^\mathrm{th}$ European Community
Framework Programme. MM acknowledges support from NASA contract NAS8-39073
and \chandra\ grant GO4-5152X.

%\bibliography{astro}

\end{document}